\documentclass[sigconf,authorversion,nonacm]{acmart}
\AtBeginDocument{%
  }

\setcopyright{acmlicensed}
\copyrightyear{2024}
\acmYear{2024}
\acmDOI{XXXXXXX.XXXXXXX}

\acmConference[VRST '24]{}{October 09--11, 2024}{Trier, Germany}
\acmISBN{978-1-4503-XXXX-X/18/06}

\begin{document}
\title{Virtual Reality for Action Evaluation}

\author{Mario De Lucas Garcia}
\email{mdelucasgarcia@hawk.iit.edu}
\authornotemark[1]
\affiliation{%
  \institution{Illinois Institute of Technology}
  \city{Chicago}
  \state{Illinois}
  \country{USA}
}

\author{Mark Roman Miller}
\email{mmiller30@iit.edu}
\authornotemark[2]
\affiliation{%
  \institution{Illinois Institute of Technology}
  \city{Chicago}
  \state{Illinois}
  \country{USA}
}

%%
%% The abstract is a short summary of the work to be presented in the article.
%% 
\begin{abstract}

Physical rehabilitation plays a crucial role in restoring functional abilities, but traditional approaches often face challenges in terms of cost, accessibility, and personalized monitoring. Asynchronous physical rehabilitation has gained traction as a cost-effective and convenient alternative, but it lacks real-time monitoring and assessment capabilities. This study investigates the feasibility of using low-cost Virtual Reality (VR) devices for action evaluation in rehabilitation exercises. We leverage state-of-the-art deep learning models and evaluate their performance on three data streams (head and hands) derived from existing rehabilitation datasets that approximate VR headset and hand data. Our results demonstrate that VR tracking data can be effectively utilized for action evaluation, paving the way for more accessible and affordable remote monitoring solutions in physical therapy. By leveraging artificial intelligence techniques and consumer-grade virtual reality technology, this study proposes an approach that could potentially address some of the challenges in asynchronous rehabilitation, such as the need for expensive motion capture systems or in-person sessions.

\end{abstract}





%% This command processes the author and affiliation and title information and builds the first part of the formatted document.
\maketitle

\section{Introduction}

In the United States, there are over 250,000 physical therapists, and this number is expected to grow by 15\% in the coming years to meet the increasing demands of society\footnote{
\url{https://www.bls.gov/ooh/healthcare/physical-therapists.htm}}.

In-person physical rehabilitation with a professional has long been the standard for therapy intervention and is widely recognized as one of the most effective ways to rehabilitate patients. However, this traditional process involves several challenges, including the need for synchronous sessions, significant time commitments, high costs, and the inconvenience of clinic travel, which may be particularly difficult for some patients \cite{machlin2011determinants}.

To address these issues, healthcare providers are increasingly adopting asynchronous rehabilitation modalities. In this model, patients perform a prescribed set of exercises at home, following guidelines provided by professionals. This approach has proven effective in musculoskeletal practices and has led to high patient satisfaction \cite{grona2018use}. However, monitoring these exercises and providing real-time assessments remain challenging tasks for researchers and companies.

% First, various depth sensors have been utilized for skeleton data acquisition, including Microsoft Kinect (discontinued), Intel RealSense\footnote{\url{https://www.intelrealsense.com/}}, and Vicon. These sensors have proven to be reliable sources of information as their output closely aligns with standard industry-level systems such as stereophotogrammetric devices \cite{merriaux2017study, capecci2016accuracy}.

Researchers have employed statistical algorithms like Dynamic Time Warping (DTW) and distance measures like Euclidean and Jaccard. Some have explored machine learning techniques like Support Vector Machines (SVM) and Gaussian Mixture Models (GMMs) to classify correct and incorrect movements. More recently, deep learning architectures such as Convolutional Neural Networks (CNNs) and Recurrent Neural Networks (RNNs) have been used to model joint dependencies and assist in movement scoring \cite{sardari2023artificial}.

Additionally, Virtual Reality (VR) technology, which is widely used in vocational training, entertainment, and medical care, offers promising applications for rehabilitation. Stand-alone head-mounted display devices are becoming more affordable, user-friendly, and accessible to many users simultaneously, providing a low-cost method for motion tracking \cite{he2022manual}.

This paper aims to combine artificial intelligence techniques with VR technology to evaluate the effectiveness of these devices as cost-effective motion-tracking systems in asynchronous rehabilitation processes. Hence, we proposed a simple baseline model that outputs the average score from the training set and an eXtreme Gradient Boosting (XGBoost) machine learning model that, given a selected set of features, predicts a score. We will provide an in-depth analysis of these models in Section \ref{sec_methods}.

We summarize the contributions of this paper as follows:
\begin{itemize}
    \item \textbf{Automated Preprocessing Pipeline:} We proposed a comprehensive preprocessing pipeline that automatically splits the data into exercise repetitions, facilitating the training of various models.
    
    \item \textbf{Model Comparison Framework:} We conducted a thorough comparison of state-of-the-art models using standardized training setups. To the best of our knowledge, this is the first work aimed at democratizing model comparison in this context.
    
    \item \textbf{Novel XGBoost Model:} We implemented a novel eXtreme Gradient Boosting (XGBoost) machine learning model that performs comparably to the current state-of-the-art models when using our training setup data and parameters.
\end{itemize} 

The rest of this work is organized as follows: The section \ref{sec_related} provides an overview of related research. Section \ref{sec_methods} details the data and models used in our experiments, including the novel eXtreme Gradient Boosting (XGBoost) machine learning model. Section \ref{sec_results} presents the results, while Sections \ref{sec_discussion} and \ref{sec_future} summarize the key findings, draw conclusions, and outline potential future research directions.

\section{Related work}
\label{sec_related}
\subsection{Motion Capture Systems}

The effectiveness of physical rehabilitation programs depends on patients' adherence and the accurate execution of prescribed exercises. To ensure systematic monitoring of exercise performance, prior research has explored the use of advanced technologies such as video-based systems, wearable devices, and Inertial Measurement Units (IMUs) for measuring and tracking motion \cite{cust2019machine}.

Vision-based strategies for motion tracking can be categorized into marker-based and marker-less techniques. Marker-based techniques involve placing tags or markers on specific body parts, which allows for precise measurement of their positions in space. This approach can be useful for guiding patients through rehabilitation exercises, as it provides accurate tracking of their movements. However, marker-based systems and body-worn sensors, while accurate for motion capture, can be intrusive and disruptive to a patient's daily activities. The need to properly place and wear the sensors or markers can make these systems impractical for successful implementation of home-based rehabilitation programs, where patients are expected to perform exercises independently without direct supervision \cite{fardoun2015proceedings, das2023comparison}.

On the other hand, marker-less techniques rely solely on computer vision algorithms to track and analyze body movements without the need for any physical markers or sensors attached to the body (Microsoft Kinect). This approach can be less intrusive and more convenient for patients, but may potentially sacrifice some accuracy compared to marker-based systems \cite{viglialoro2019review, das2023comparison}.

\subsection{Applied Virtual Reality}
Virtual Reality (VR) technology has gained significant attention in recent years, with numerous research studies exploring its potential applications across various domains involving physical movement and motion activities. These applications range from learning complex skills, such as couple dancing, to enhancing physiotherapy and rehabilitation processes. By leveraging the immersive and interactive nature of VR, researchers and practitioners have been able to develop innovative approaches that show promising results in terms of improved learning outcomes, patient engagement, and overall effectiveness.

One notable example is the use of VR in learning couple dance. By creating virtual environments that simulate real-world dance settings, VR systems can provide learners with immersive experiences that help them practice and refine their dance moves. These systems can offer real-time feedback, personalized guidance, and the ability to practice at one's own pace, making the learning process more efficient and enjoyable \cite{senecal2020salsa}.

Similarly, VR technology has been applied to the field of physiotherapy. By incorporating VR into rehabilitation programs, therapists can create engaging and interactive exercises that motivate patients to adhere to their treatment plans. VR-based physiotherapy can also provide real-time monitoring and analysis of patient movements, enabling therapists to track progress, adjust treatment plans, and provide targeted feedback \cite{postolache2019tailored, chirico2020virtual, laver2017virtual}.

\subsection{Action evaluation}
In prior studies, movement evaluation is typically conducted by comparing a patient’s performance to that of healthy participants. Early researchers employed machine learning models to classify individual movements as either correct or incorrect. They experimented with well-known methods such as the Adaboost classifier, k-nearest neighbours, and Bayesian classifier, which produce binary class outputs. However, these approaches lack the ability to detect varying levels of motion quality or to identify gradual changes in patient performance throughout the rehabilitation program \cite{zhang2011template, ar2014computerized}.

A line of research has utilized probabilistic approaches for modeling and evaluating rehabilitation actions. Studies employing hidden Markov models (HMMs) \cite{capecci2018hidden} and Gaussian Mixture Models (GMMs) \cite{lin2016movement} generally conduct quality assessments based on the likelihood that individual movement sequences are drawn from a trained model.

In recent years, researchers have been exploring the application of deep learning techniques to model complex dependencies between body joints for the purpose of assessing physical exercises. One notable study by Liao et al. \cite{liao2020deep} introduced the Spatio-Temporal Neural Network (STNN), a deep learning architecture designed specifically for evaluating the quality of exercise performance. The STNN combines several key components, including temporal pyramids, multi-branch convolution, and recurrent layers, to effectively capture and analyze the spatial and temporal patterns in human motion data. The input to the STNN is typically skeleton data, which represents the positions and movements of 22 to 39 body joints, depending on the sensors used to capture the motion.

Another important contribution in this field comes from Deb et al. \cite{deb2022graph}, who employed Spatio-Temporal Graph Convolution Networks (SGNN) for the task of physical exercise assessment. SGNNs are a type of deep learning model that can operate directly on graph-structured data, such as the skeletal representation of human body joints. By leveraging the inherent graph structure of the body, SGNNs can effectively capture the dependencies and relationships between different joints, enabling a more accurate and robust assessment of exercise performance.

In a different approach, Mottagui et al. \cite{mottaghi2022automatic} proposed a hybrid framework that combines deep learning algorithms with probabilistic models. By integrating these two complementary paradigms into a unified training process, their method aims to leverage the strengths of both deep learning and probabilistic modelling. Deep learning excels at automatically learning complex patterns and representations from raw data, while probabilistic models provide a principled way to handle uncertainty and incorporate prior knowledge. By combining these approaches, Mottagui et al. \cite{mottaghi2022automatic} seek to develop more robust and interpretable models for exercise assessment.

These studies highlight the growing interest in applying advanced machine learning techniques, particularly deep learning, to the problem of modelling and assessing complex human movements in the context of physical exercises.

\section{Methods}
\label{sec_methods}
\subsection{Problem formulation}

For each subject $s \in S$ the set of all subjects listed in the dataset, we denote her joint positions and orientations for $i$th exercise (i.e. each exercise file may have more than one repetition) as $P_{s, i}$ and $O_{s, i}$, respectively. Each exercise file has a ground-truth score annotation $y_{s, i} \in [0, 1]$ that represents the quality of the performed exercise based on professional criteria. The higher the score, the better the patient moves. We train each model $\text{Model}_i$ for each $i$ distinct exercise with its $\theta_i$ parameters.

Each model we have tested predicts a continuous score $\hat{y}_{j}$ close to the ground-truth assessment score, $y^*_{j}$ for a given joint position or orientation. 

\begin{equation}
    \hat{y} = \text{Model}_i (P_{S, N} ; \theta_i), \text{       s.t.    } \hat{y}_{j} \approx y^*_{j}
\end{equation}
or using joint orientations as inputs, as follows
\begin{equation}
    \hat{y} = \text{Model}_i (O_{S, N} ; \theta_i), \text{       s.t.    } \hat{y}_{j} \approx y^*_{j}
\end{equation}

\subsection{Action Evaluation Public Datasets}
The requirements we established for useful datasets were: (1) it must be publicly available; (2) it must have a clearly defined exercise quality measure; and (3) it must contain skeleton data, specifically including head and hands data. Consequently, we thoroughly researched public options for evaluated action data. Our findings indicate that KIMORE is the most reliable data source, as it meets all our requirements. Table \ref{tab_datasets} provides a comparison of each option.

\begin{table*}[]
\caption{Publicly available datasets.}
\label{tab_datasets}
\begin{tabular}{ p{0.1\linewidth}p{0.25\linewidth}p{0.35\linewidth}p{0.05\linewidth}p{0.1\linewidth} }
\toprule
\textbf{Dataset Name}  & \textbf{Exercise Quality Measure?}                                                                                                          & \textbf{Position and Rotation Data?}                                                                                                                                                                         & \textbf{Access?} & \textbf{Notes}                                                           \\
\midrule
KIMORE                 & Yes - rated by a physician.                                                                                                                 & Yes. Skeleton positions (trajectory positions) and orientations in the format of Kinect.                                                                                                                     & Yes              & 78 subjects                                                              \\
IntelliRehabDS (IRDS)  & No, the dataset contains a correctness label for each gesture, indicating whether it was performed correctly or incorrectly by the subject. & Yes, the dataset contains the 3D coordinates of 25 body joints extracted by a Kinect sensor, as well as the raw depth map images for each frame.                                                             & Yes              & 29 subjects, out of which 15 were patients and 14 were healthy controls. \\
UI-PRMD                & No, the paper proposes a taxonomy for performance metrics for the evaluation of therapy movements, but only a reduced subset was scored.    & Yes, the paper provides both position and angle data for 39 joints of the human body, obtained from a Vicon optical tracker and a Kinect sensor.                                                             & Yes              & 10 people, 10 actions                                                    \\
3D Motion Capture Data & No                                                                                                                                          & Yes, the paper provides 3D joint centre positional data, 3D joint angles, and 3D segment velocity and acceleration data of the head, trunk, upper arms, forearms, pelvis, thighs, shanks, and feet segments. & No              & 183 Athletes                                                            
\\
\bottomrule
\end{tabular}
\end{table*}

For running the experiments, we used the KIMORE dataset, which contains data from 78 subjects: 44 in the control group (i.e., experts or non-experts who are healthy) and 34 with chronic motor disabilities. This dataset was recorded using the Kinect v2 depth sensor, capturing 25 distinct body parts, including their skeleton positions and orientations. Each subject performed five exercises:

\begin{itemize}
    \item Lifting of the arms
    \item Lateral tilt of the trunk with the arms extended
    \item Trunk rotation
    \item Pelvis rotations on the transverse plane
    \item Deep squats
\end{itemize}

\begin{figure}[h]
  \centering
  \includegraphics[width=\linewidth]{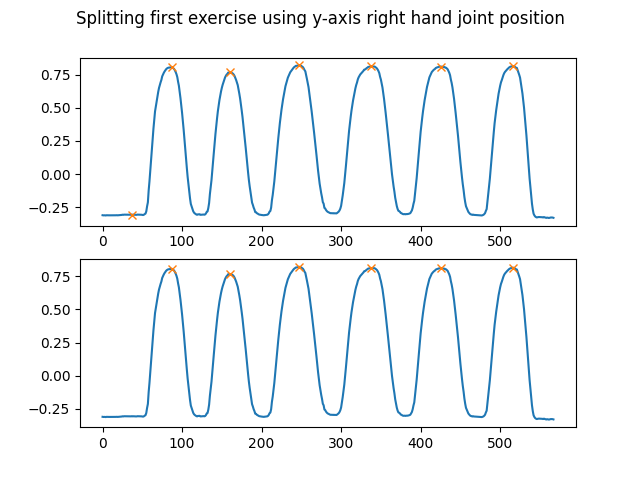}
  \caption{The above image shows all the peaks it found. The below image shows the final start timestamps we used for splitting the CSV file into repetitions.}
  \Description{A comparison between the peaks it found and the actual values we used for the repetitions.}
  \label{fig_ex1preprocessing}
\end{figure}

During preprocessing, as described in the original paper, we applied a 3rd order low-pass Butterworth filter to remove temporary spikes. The dataset consists of a single CSV file containing positions or orientations per person and per exercise. We developed a custom function to split these files into individual repetitions. For example, in the lifting of the arms exercise, we identified that the Y coordinate of hands effectively indicated when the patient started a repetition. Additionally, we applied threshold filtering to remove false peaks that were found. We used the timestamps where the X position was maximal to mark the start of each repetition. See Figure \ref{fig_ex1preprocessing}

However, a limitation of the KIMORE dataset is that it provides only one score per exercise, meaning all repetitions share the same ground truth score. The given score ranges from 0 to 50, which we scaled to a range of 0 to 1 for our analysis.

\subsection{Proposed Method}
We have built two baseline models to compare the precision of current state-of-the-art models for the KIMORE dataset. These baseline models serve as a reference point for evaluating the performance of more advanced techniques.

The first baseline model is an aggregator that always outputs the average score of the training repetitions. Since the dataset contains a variety of different exercises, we apply this baseline model to each exercise separately. 
% This model provides a simple and straightforward approach to establishing a baseline performance.

The second baseline model is an eXtreme Gradient Boosting (XGBoosting) machine learning model that utilizes a curated list of features. Our goal is to determine whether Virtual Reality data can be effectively used to assess physical rehabilitation exercises. To this end, we calculate the following 44 features:

\begin{itemize}
    \item Maximum, minimum, mean, and standard deviation of the distance between hands for the three coordinates.
    \item Maximum, minimum, mean, and standard deviation of the distance between the right/left hand and the head for the three coordinates.
    \item Maximum, minimum, mean, and standard deviation for each right/left hand and head joint for each coordinate.
\end{itemize}

Our proposed model demonstrates competitive performance, even outperforming state-of-the-art models like the Spatio-Temporal Neural Network (STNN) in various exercises from the KIMORE dataset (see Figure \ref{res_ori_full_ex1}).

\section{Results}
\label{sec_results}

\subsection{Setup}
We first conducted a thorough review of publicly available datasets and chose to work with the KIMORE dataset. Although UI-PRMD and IntelliRehabDS include a greater variety of exercises, they lack a clear methodology for scoring. Specifically, UI-PRMD provides a labeled subset with scores derived from distance functions, while IntelliRehabDS only offers binary labels (i.e., correct or incorrect movements). The KIMORE dataset, by contrast, provides a more comprehensive and methodologically sound framework for evaluating rehabilitation exercises, making it the most suitable choice for our study.

Next, we developed a preprocessing pipeline to segment each recording into individual exercise repetitions. This was accomplished by detecting peaks in the position data of key joints. We then applied a cleaning filter to eliminate false peaks. To ensure generality, we standardized each repetition to a fixed length of 104 timesteps. Additionally, we prepared the data so that any model could be trained on either the whole body joints or just the head and hand joints.

Then, we evaluate the performance of each model using three metrics where each model was trained for each exercise separately since the inherent features of each movement were disparate.: (1) Root Mean Square Error (RMSE), (2) Mean Absolute Error (MAE), and (3) Mean Absolute Percentage Error (MAPE). Lower scores indicate more precise predictions across all metrics. We use the same training and testing subsets for each model to ensure a fair comparison and eliminate the influence of external factors. Additionally, we modify each model to accept either the whole body joint positions and orientations or only the head and hand data.

We maintain the training procedures described in the original papers for the Spatio-Temporal Neural Network (STNN) and Spatio-Temporal Graph Convolution Network (SGNN) models. Specifically, we train a separate STNN for each exercise, using 500 epochs, a batch size of 10, and a learning rate of 0.0001. Similarly, we train a separate GCN for each exercise, using 1000 epochs, a batch size of 10 and a learning rate of 0.0001.

\subsection{Position}
The results of training the four models with the same data are presented below. In this context, \textit{\textbf{VR}} indicates that only head and hand data were used.

\begin{figure*}[h]
  \centering
  \includegraphics[width=\linewidth]{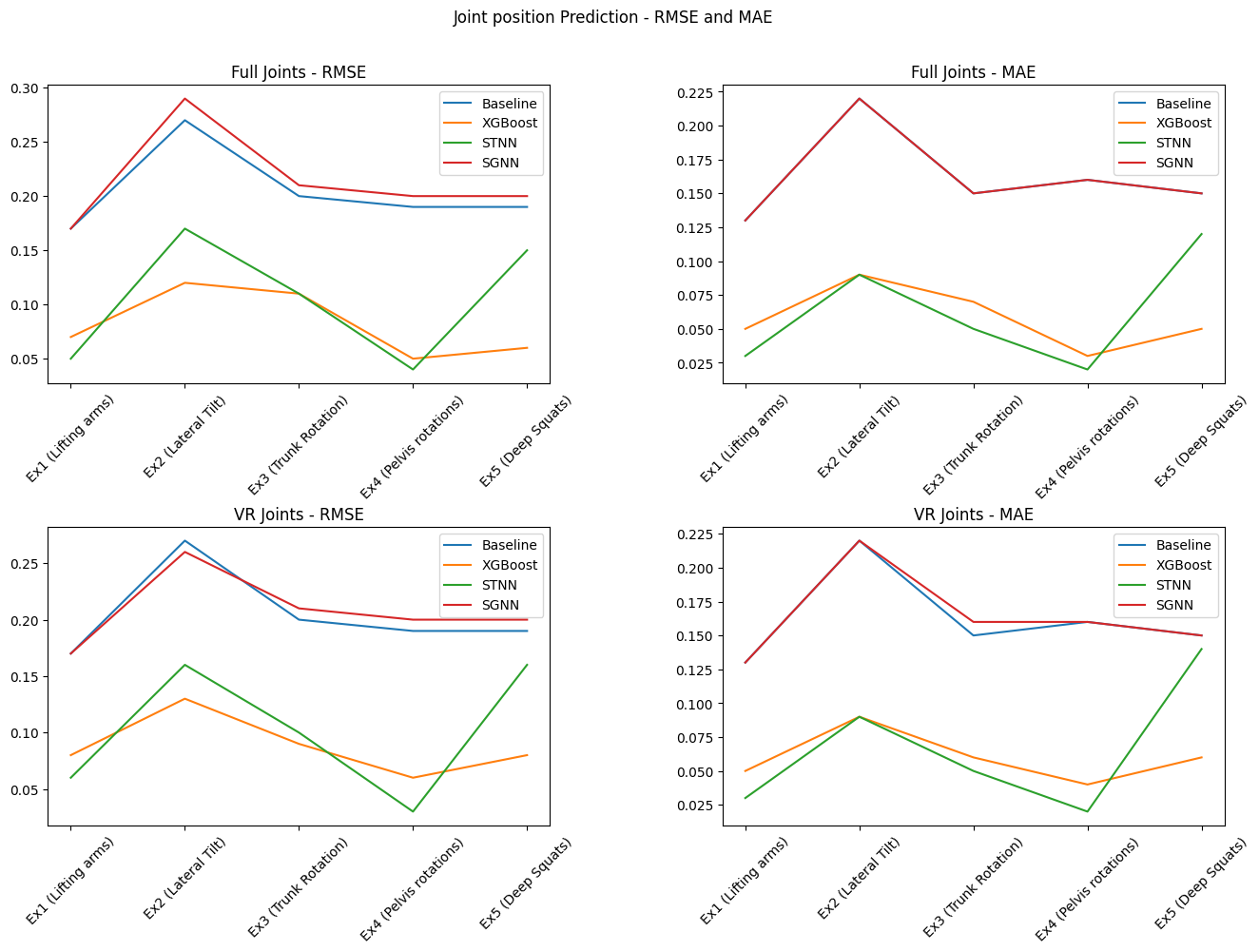}
  \caption{Results of training the four models with the same data. VR means only head and hand data.}
  \Description{It plots four different diagrams showing the RMSE and MAE.}
  \label{fig_pos_results}
\end{figure*}

As expected, the Baseline model consistently performs the worst across all exercises, both for the full joint data and the VR (head and hand) data (See Figure \ref{fig_pos_results}). This simple averaging approach serves as a reasonable lower bound but is outperformed by more sophisticated models. 

Surprisingly, according to the full joints chart, the SGNN model exhibits a worse RMSE compared to the simple baseline model and has an MAE that is equally poor. In contrast, the XGBoost model outperformed the STNN model in exercises 3 and 5, and it matched the STNN model in exercise 4, based on the RMSE metric. When evaluating the MAE metric, XGBoost generally performed worse than the STNN model, except for the last exercise. 

This trend persisted when using VR data format (head and hand data only). Interestingly, both models maintained their performance levels even when limited to head and hand joint positions. Notably, the XGBoost model showed improved performance in the third exercise under these conditions. See Figure \ref{fig_pos_results}.

\subsection{Orientation}
\begin{figure*}[h]
  \centering
  \includegraphics[width=\linewidth]{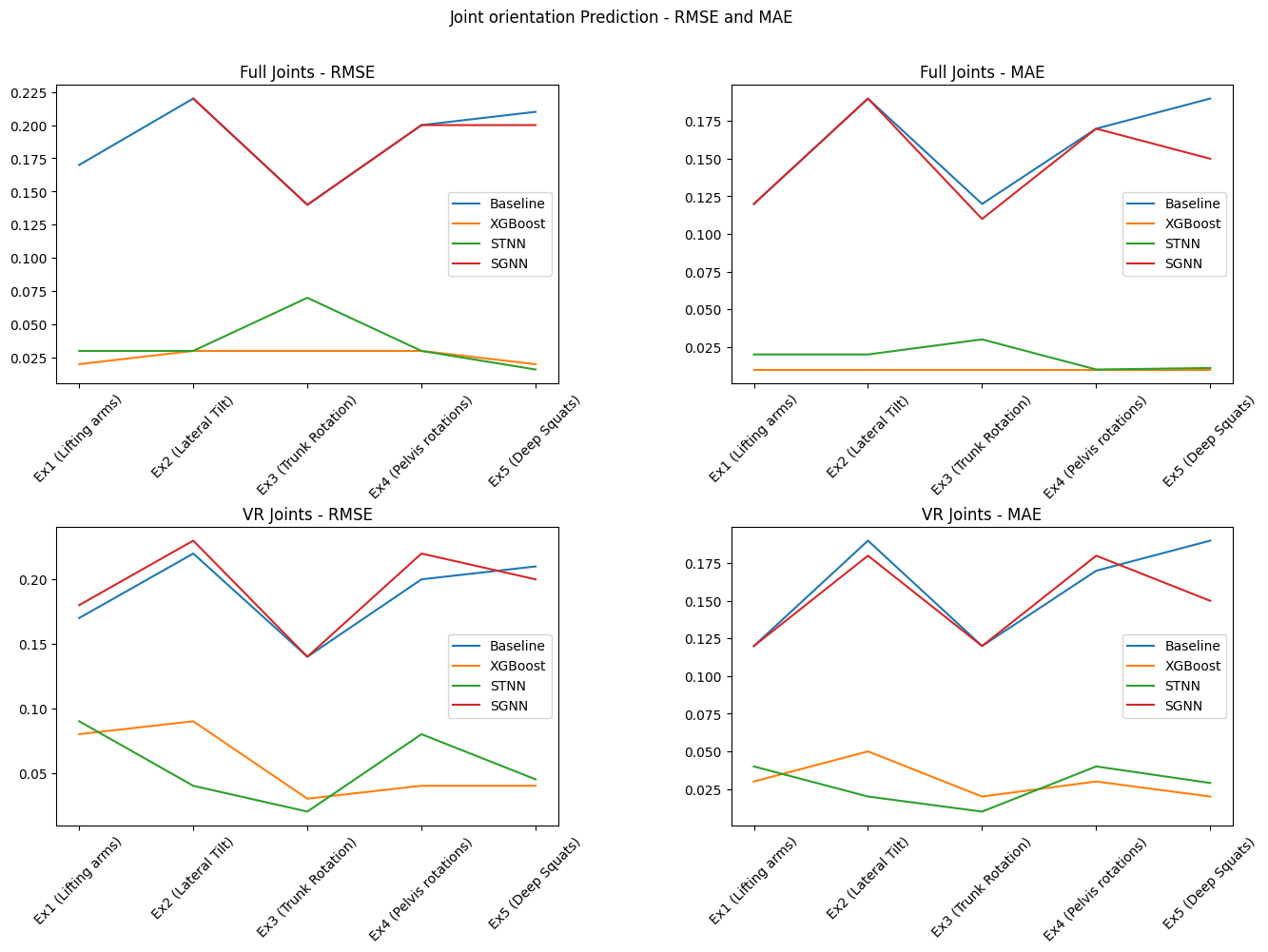}
  \caption{Results of training the four models with the same data. VR means only head and hand data.}
  \Description{It plots four different diagrams showing the RMSE and MAE.}
  \label{fig_ori_results}
\end{figure*}

As mentioned earlier, the baseline model consistently exhibits the poorest performance across all workout scenarios, both when considering the full joint data and the virtual reality (VR) data comprising head and hand orientations. Similarly, the SGNN (Spatial-Temporal Graph Neural Network) model mirrors the unsatisfactory performance of the baseline model, as illustrated in Figure \ref{fig_ori_results}.

In contrast, the XGBoost model closely reaches the performance of the STNN (Spatial Temporal Neural Network) model for most exercises. This tendency continues even when employing the VR data format. Nonetheless, both the XGBoost and STNN models experience a significant decline in performance when working with limited joint orientation data, as evidenced by Figure \ref{fig_ori_results}.

\subsection{Position VS Orientation}
\begin{figure*}[h]
  \centering
  \includegraphics[width=\linewidth]{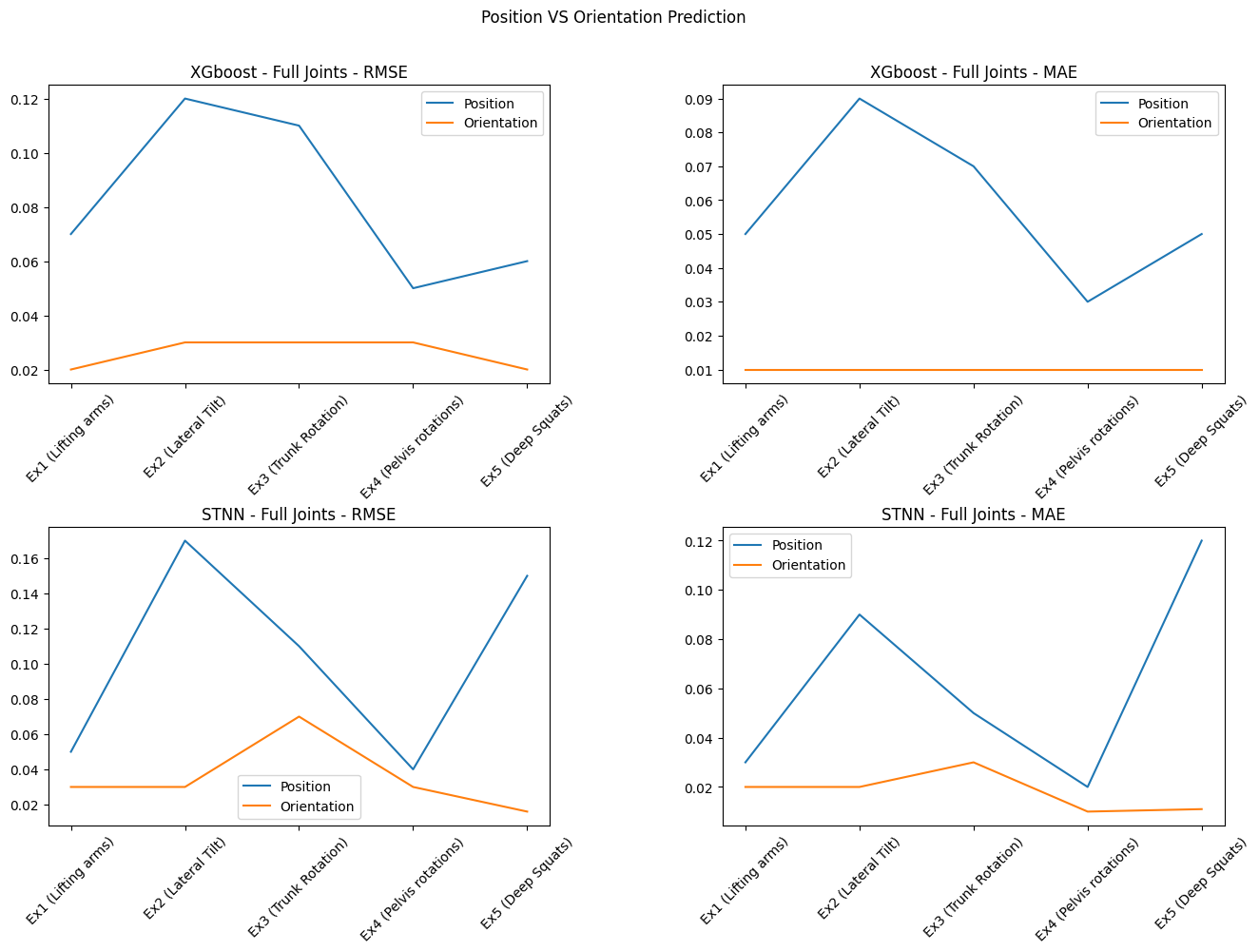}
  \caption{Comparison between the XGBoost and STNN models using the full joint positions and orientations.}
  \Description{It compares the performance between XGBoost and STNN}
  \label{fig_pos_ori}
\end{figure*}

To optimize performance, it is essential to determine which data input format yields superior results. Focusing on the full joint data analysis, as depicted in Figure \ref{fig_pos_ori}, we observe that the Spatio-Temporal Neural Network (STNN) model consistently outperforms across all exercise routines when leveraging orientation data as input compared to when using position data alone. This behavior aligns with the expected outcome, as the STNN architecture was specifically designed to capitalize on orientation data, considering the spatial relationships and orientations of the body joints during movement.

Interestingly, our findings also reveal that the XGBoost model, a powerful ensemble learning technique, performs significantly better when incorporating orientation data as an additional input feature. This suggests that the inclusion of joint orientation information, in addition to positional data, provides valuable insights that enable the XGBoost model to make more accurate predictions and classifications across exercise routines.

\begin{figure*}[h]
  \centering
  \includegraphics[width=\linewidth]{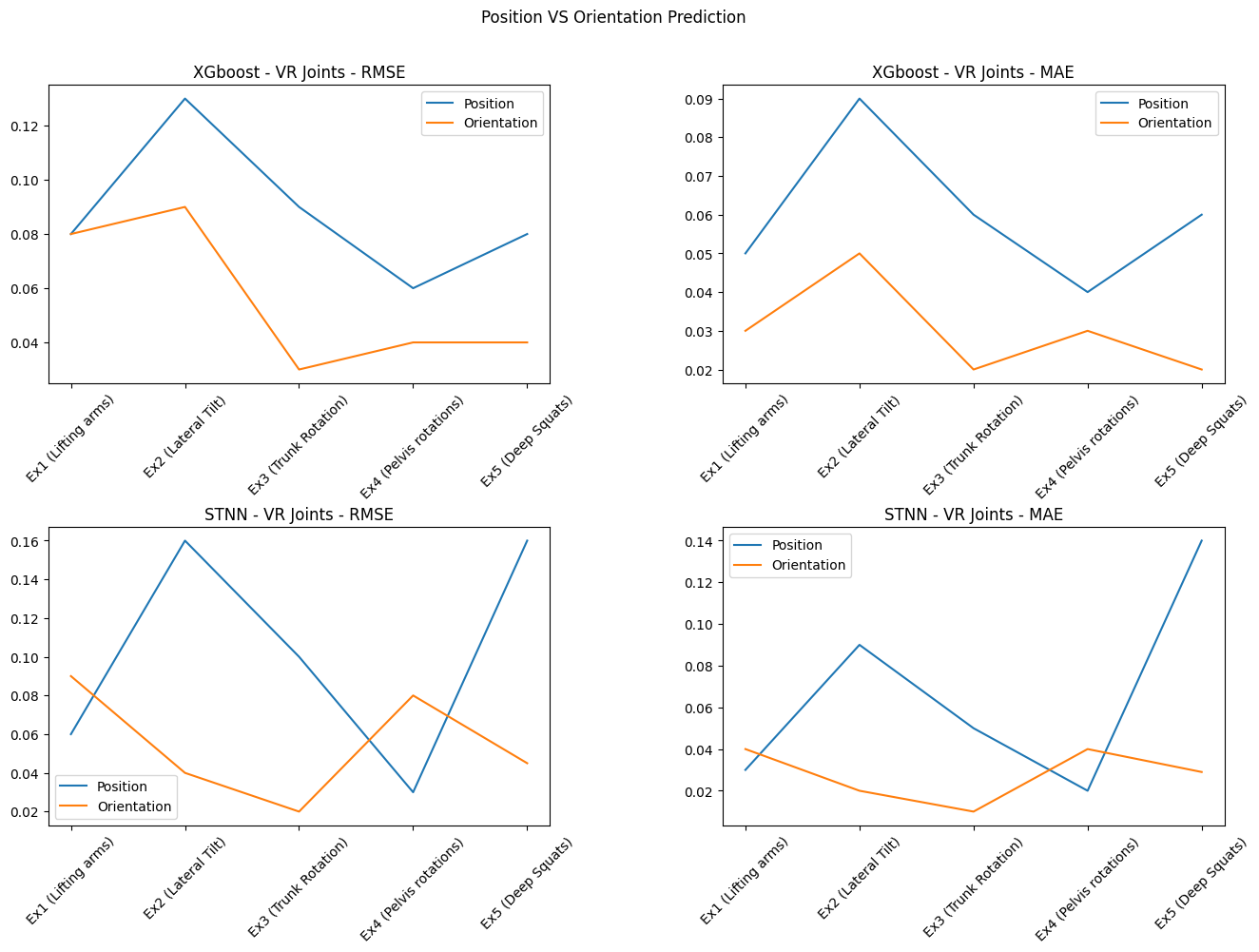}
  \caption{Comparison between the XGBoost and STNN models using the head and hand joint positions and orientations.}
  \Description{It compares the performance between XGBoost and STNN}
  \label{fig_vr_pos_ori}
\end{figure*}

Shifting our focus to the head and hand data analysis, as illustrated in Figure \ref{fig_vr_pos_ori}, we observe that the trend discovered in the full joint data analysis persists. Specifically, the Spatio-Temporal Neural Network (STNN) model continues to exhibit superior performance when leveraging orientation data as input compared to position data alone. However, it is essential to note that exercise 4, which involves pelvis rotations on the transverse plane, stands as an exception to this trend. For this particular exercise, the STNN model surprisingly achieved better results when working with position data rather than orientation data.

This discovery is intriguing because prior research studies in this domain did not explicitly explore the implications of employing either position or orientation data as input features for their respective analyses. Most previous works focused on leveraging a specific data format without directly comparing the performance implications of using alternative input representations, such as position versus orientation information.

\section{Discussion}
\label{sec_discussion}
Advances in machine learning and computer vision techniques have sparked increased interest in the automated evaluation of rehabilitation exercises. However, despite significant progress, several open questions and numerous challenges remain before these asynchronous systems can be widely deployed.

To encourage adherence to rehabilitation programs, it is crucial to design real-time assessment applications that guide patients through each exercise. However, this remains a notable challenge.

In this work, we present a novel approach for handling skeleton data using an ensemble learning method. Specifically, we developed an eXtreme Gradient Boosting (XGBoost) model that achieves performance comparable to state-of-the-art models on the KIMORE dataset. Additionally, we provide a common framework that allows researchers to easily integrate and evaluate new models against existing ones using our open preprocessing and splitting pipeline.

Initially, we analyzed various public datasets relevant to this problem. After a thorough comparison, we decided to focus exclusively on the KIMORE dataset. We developed an automatic approach to segment the data into exercise repetitions by detecting the peaks in certain joint movements.

Our results indicate that the Spatio-Temporal Neural Network (STNN) generally outperforms other models, even surpassing the results reported by its original authors. However, as shown in Figures \ref{fig_pos_ori} and \ref{fig_vr_pos_ori}, our XGBoost model achieves performance comparable to the STNN across all exercises. This finding suggests that machine learning models can remain competitive even as more complex deep learning models emerge. Sometimes, a simpler yet well-tuned approach can be more effective than a more intricate solution.

Furthermore, we observed that using only head and hand data results in a slight performance drop but still remains close to state-of-the-art values. This suggests that head-mounted virtual reality devices could potentially be used in rehabilitation programs as a low-cost method for tracking and assessing exercises.

All the code will be available from \url{https://github.com/mdelucasg/VR-for-Action-Evaluation}.

\section{Future work}
\label{sec_future}
To further validate and corroborate our findings, we have planned to conduct additional data collection using head-mounted virtual reality (VR) devices. While our current analysis has demonstrated that, in theory, even with a loss in performance, the use of limited joint orientation data remains relatively affordable and feasible, it is crucial to verify these observations with real-world data gathered from actual VR devices. This step is essential to fully verify our claims and ensure the robustness of our conclusions.

Furthermore, we intend to revisit the implementation of the Spatio-Temporal Graph Neural Network (STGNN) model. Despite its promising theoretical foundations, the performance we have observed from this model in our current analysis falls short of state-of-the-art (SOTA) levels. By re-implementing and training the SGNN architecture, we aim to unlock its full potential and push its performance closer to the cutting edge in this domain.

% The ``\verb|acmart|'' document class includes the ``\verb|booktabs|''
% package --- \url{https://ctan.org/pkg/booktabs} --- for preparing
% high-quality tables.

% \begin{table*}
%   \caption{Some Typical Commands}
%   \label{tab:commands}
%   \begin{tabular}{ccl}
%     \toprule
%     Command &A Number & Comments\\
%     \midrule
%     \texttt{{\char'134}author} & 100& Author \\
%     \texttt{{\char'134}table}& 300 & For tables\\
%     \texttt{{\char'134}table*}& 400& For wider tables\\
%     \bottomrule
%   \end{tabular}
% \end{table*}

% \begin{figure}[h]
%   \centering
%   \includegraphics[width=\linewidth]{sample-franklin}
%   \caption{1907 Franklin Model D roadster. Photograph by Harris \&
%     Ewing, Inc. [Public domain], via Wikimedia
%     Commons. (\url{https://goo.gl/VLCRBB}).}
%   \Description{A woman and a girl in white dresses sit in an open car.}
% \end{figure}

\bibliographystyle{ACM-Reference-Format}
\bibliography{references}

% \appendix

% \section{Preprocessing}

\end{document}